\documentclass[aps,prb,superscriptaddress,twocolumn]{revtex4}
\usepackage{amsmath}
\usepackage{amssymb}
\usepackage{graphicx}

\begin{document}

\title{Persistence of the 0.7 anomaly of quantum point contacts in high magnetic fields}

\author{E.~J.~Koop}
\email[e-mail: ]{e.j.koop@rug.nl}
\author{A.~I.~Lerescu}
\author{J.~Liu}
\author{B.~J.~van~Wees}
\affiliation{Physics of Nanodevices Group, Zernike Institute for Advanced Materials,\\
University of Groningen, Nijenborgh 4, 9747 AG  Groningen, The Netherlands}
\author{D.~Reuter}
\author{A.~D.~Wieck}
\affiliation{Angewandte Festk\"{o}rperphysik, Ruhr-Universit\"{a}t Bochum, D-44780 Bochum,
Germany}
\author{C.~H.~van~der~Wal}
\affiliation{Physics of Nanodevices Group, Zernike Institute for Advanced Materials,\\
University of Groningen, Nijenborgh 4, 9747 AG  Groningen, The
Netherlands}

\date{\today}


\begin{abstract}

The spin degeneracy of the lowest subband that carries
one-dimensional electron transport in quantum point contacts appears
to be spontaneously lifted in zero magnetic field due to a
phenomenon that is known as the 0.7 anomaly. We measured this energy
splitting, and studied how it evolves into a splitting that is the
sum of the Zeeman effect and a field-independent exchange
contribution when applying a magnetic field. While this exchange
contribution shows sample-to-sample fluctuations, it is for all QPCs
correlated with the zero-field splitting of the 0.7 anomaly. This
provides evidence that the splitting of the 0.7 anomaly is dominated
by this field-independent exchange splitting.

\end{abstract}

\maketitle


A quantum point contact (QPC) is a short ballistic transport channel
between two electron reservoirs. Its conductance as a function of
channel width is quantized\cite{vanWees1988,Wharam1988} and shows
plateaus at integer multiples of $2e^2/h$, where $e$ the electron
charge and $h$ Planck's constant. Remarkably, almost all
semiconductor QPCs show an additional plateau at $\sim 0.7(2e^2/h)$.
Earlier
work\cite{Thomas1996,Thomas1998,Kristensen2000,Cronenwett2002,Starikov2003,Reilly2005,DiCarlo2006,Rejec2006}
related this "0.7 anomaly" to a spontaneous removal of spin
degeneracy in zero magnetic field for the lowest one-dimensional
subband in the QPC.

\begin{figure}[h!]
  \includegraphics[width=0.78\columnwidth]{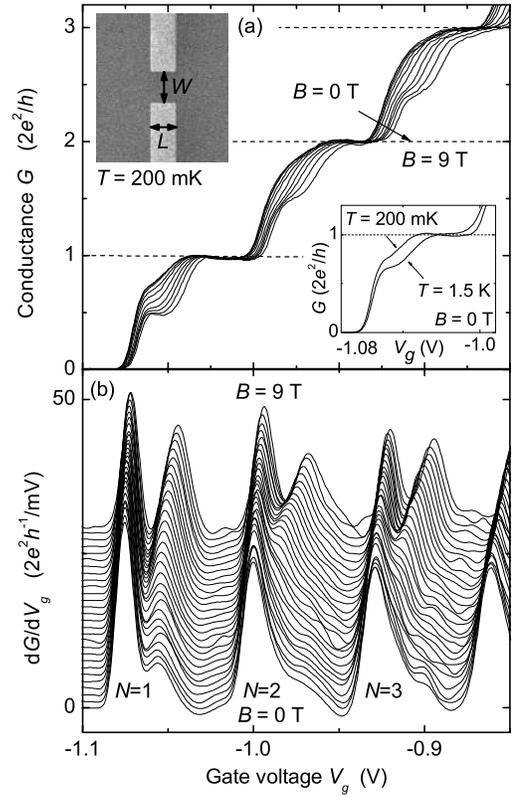}\\
  \caption{{\bf Conductance of a QPC. a,} The differential conductance $G$ as a function of
  gate voltage $V_g$ at 200 mK,
  for a QPC with $L$ = 300 nm and $W$ = 400 nm. The in-plane magnetic field is
  increased from $B = 0 \; {\rm T}$ to $B = 9 \; {\rm T}$. The first three spin-degenerate
  plateaus at integer multiples of $2e^2/h$ for $B = 0 \; {\rm T}$ split into six spin-resolved plateaus
  integer multiples of $e^2/h$ for $B = 9 \; {\rm T}$.
  The right inset zooms in on the 0.7 anomaly in zero field, and how it is enhanced as the temperature
  is increased.
  The left inset shows a micrograph of a split gate QPC of
  length $L$ and width $W$. {\bf b,} Transconductance ${\rm d}G / {\rm d}V_g$ traces
  (offset vertically for clarity) obtained by
  differentiating the data in ({\bf a}). The 0.7
  anomaly appears as a splitting of the transconductance peak for the $N=1$ subband for $B = 0 \; {\rm T}$.}
  \label{FIG:ConductanceVSVgateFieldDep}
\end{figure}

Understanding the 0.7 anomaly has been the topic of on-going
research for more than a decade now
\cite{Fitzgerald2002,Ennslin2006}, and is of interest for
spintronics and quantum information proposals where QPCs are a key
element. It cannot be described with a single-particle theory, and
QPCs are therefore also an important model system for studies of
many-body physics in nanodevices.  Recent progress includes evidence
that the 0.7 anomaly has similarities with transport through a Kondo
impurity \cite{Cronenwett2002,Meir2002}. This is now supported by
numerical simulations of realistic QPC geometries that show that a
bound spin-polarized many-body state can spontaneously form inside
an open QPC system \cite{Rejec2006}. Graham \textit{et al.} reported
evidence that many-body effects also play a role in magnetic fields
at crossings between Zeeman levels of different subbands
\cite{Graham2003}, and at crossings of spin-split subbands with
reservoir levels \cite{Graham2005}. We report here how the many-body
effects in QPCs depend on the QPC geometry. We study spin-splittings
within one-dimensional subbands, both in zero field and high
magnetic fields. While this type of data from individual devices has
been reported before \cite{Thomas1998,Cronenwett2001}, we report
here data from a set of 12 QPCs with identical material parameters.
This allows us to draw attention to an exchange contribution to the
splitting in high magnetic fields, and to correlate this with the
physics that underlies the 0.7 anomaly.

We realized our QPCs using a GaAs-AlGaAs heterostructure with a
two-dimensional electron gas (2DEG) 114 nm below its surface (see
Methods). A QPC is formed by applying a negative gate voltage
$V_{g}$ to a pair of electrodes on the wafer surface. The 2DEG below
the electrodes is then fully depleted, and tuning of $V_g$ allows
for controlling the width of a short one-dimensional transport
channel. Our QPCs had different values for the length $L$ and width
$W$ for the electrode spacing that defines the device (inset of
Fig.~\ref{FIG:ConductanceVSVgateFieldDep}a). Note that $W$ should
not be confused with the actual width of the transport channel that
is controlled with $V_{g}$.

Figure~\ref{FIG:ConductanceVSVgateFieldDep}a presents the
differential conductance $G$ of a QPC as a function of $V_{g}$, with
the source-drain voltage $V_{sd} \approx 0$. Increasing $V_{g}$ from
pinch-off ($G=0$) lowers and widens the saddle-point-like potential
that defines the short transport channel. Consequently, an
increasing number of one-dimensional subbands gets energies below
the Fermi level. In zero magnetic field, this results in a step of
$2e^2/h$ in the conductance each time an additional subband starts
to contribute to transport. We label these spin-degenerate subbands
with a number $N$, starting with $N=1$ for the lowest subband. With
a high in-plane magnetic field $B$ the spin degeneracy within each
subband $N=1,2,3...$ is lifted, and the conductance increases now in
steps of $e^2/h$.

\begin{figure}[h!]
  \includegraphics[width=1\columnwidth]{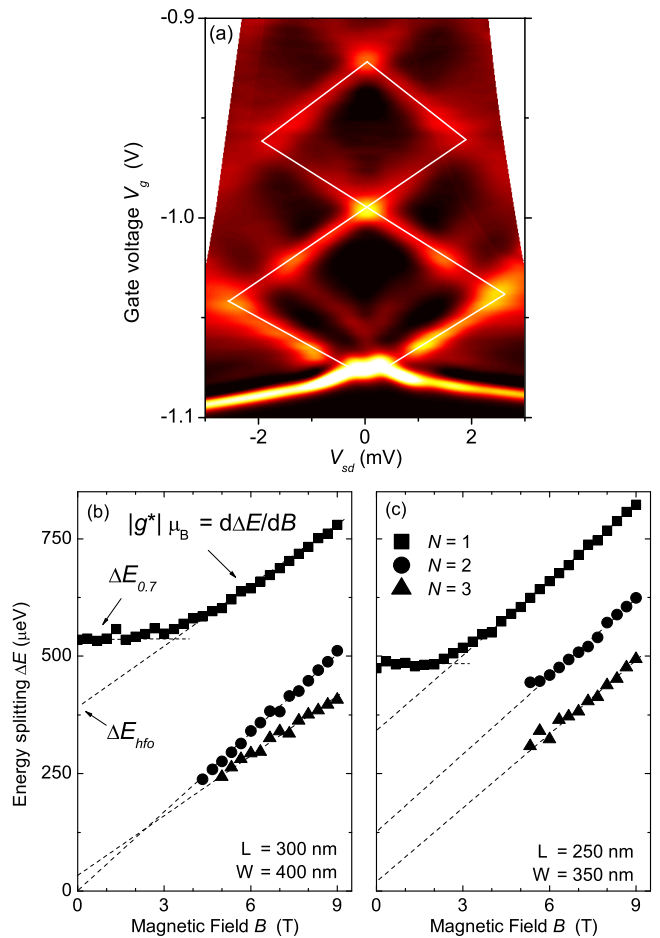}\\
  \caption{{\bf Energy splittings as a function of magnetic field. a,} Colorscale plot
  with transconductance ${\rm d}G / {\rm d}V_g$ as a
  function of source-drain voltage $V_{sd}$ and $V_g$ at $B$ = 0 and $T$ = 200 mK,
  for a QPC with $L$ = 300 nm and $W$ = 400 nm. Dark
  regions represent plateaus in the differential conductance,
  where ${\rm d}G / {\rm d}V_g \approx 0$.
  The color scale via red towards white represent peaks in transconductance.
  White lines are added as a guide to the eye and mark transitions
  between various conductance plateaus.
  {\bf b, c,} Energy splittings $\Delta E$ as obtained from
  transconductance ${\rm d}G / {\rm d}V_g$ traces,
  as a function of magnetic field at
  200~mK, for a QPC with $L$ = 300 nm and $W$ = 400 nm and a QPC with $L$ = 250 nm and $W$ = 350 nm.
  The traces present splittings for the subbands $N=1,2,3$, see the legend in ({\bf c}).
  These data sets for $\Delta E$ are characterized (results presented in Fig.~\ref{FIG:GfactorAndEnergySplittingVSShape})
  with three parameters for each subband $N=1,2,3$.
  The slope
  of $\Delta E$ versus $B$ in the high fields ($B > 5 \; T$) is
  characterized with the effective g-factor $|g^*|$.
  A linear fit on $\Delta E$ from this regime, extrapolated down
  to zero field, shows and offset from a linear Zeeman effect, characterized by the
  high-field offset $\Delta E_{hfo}$.
  At low field $\Delta E$ for $N=1$ saturates at the
  energy splitting of the 0.7 anomaly, which is characterized by $\Delta E_{0.7}$.
  In the low-field regime, $\Delta E$ values could not be obtained for $N=2$ and $N=3$.
  The high-field data shows a decrease in the offset $\Delta E_{hfo}$ with increasing subband index $N$.}
  \label{FIG:EnergySplittingVSMagneticField}
\end{figure}

We use this type of data to determine the energy splitting $\Delta
E$ between spin-up and spin-down levels within the subbands
$N=1,2,3$. The onset of transport through a next (spin-polarized)
subband appears as a peak in transconductance (${\rm d}G / {\rm
d}V_g$) traces as in Fig.~\ref{FIG:ConductanceVSVgateFieldDep}b,
which we derive from traces as in
Fig.~\ref{FIG:ConductanceVSVgateFieldDep}a. We determine the peak
spacing $\Delta V_{g}$ along the $V_{g}$ axis within each subband by
fitting two peaks per subband on the transconductance traces (see
Methods). Subsequently, transconductance data from non-linear
transport measurements is used for converting $\Delta V_{g}$ values
into energy splittings $\Delta E$
(Fig.~\ref{FIG:EnergySplittingVSMagneticField}a). Here, the onsets
of conductance plateaus appear as diamond shaped patterns in the
$V_{sd} - V_{g}$ plane. The width of these diamonds along the
$V_{sd}$ axis defines the subband spacing (a measure for the degree
of transverse confinement in the channel), and we use this to
determine the spacing $\hbar \omega_{12}$ between the $N=1$ and
$N=2$ subband. The slopes of the diamonds can be used to convert a
gate-voltage scale into energy scale \cite{Patel1991}.

The 0.7 anomaly is clearly visible in the data set presented in
Fig.~\ref{FIG:ConductanceVSVgateFieldDep}. The conductance trace for
zero field in Fig.~\ref{FIG:ConductanceVSVgateFieldDep}a shows
besides pronounced steps of $2e^2/h$ an additional shoulder at $\sim
0.7(2e^2/h)$, which becomes more pronounced at higher temperatures
(inset of Fig.~\ref{FIG:ConductanceVSVgateFieldDep}a). With
increasing magnetic field, the 0.7 anomaly evolves into the first
spin-resolved conductance plateau at $e^2/h$. In
Fig.~\ref{FIG:ConductanceVSVgateFieldDep}b the 0.7 anomaly appears
as a zero-field splitting in the transconductance peak for $N=1$,
which evolves into two spin-split peaks in high fields. In earlier
work this observation was the basis for assuming that the 0.7
anomaly results from a spontaneous removal of spin degeneracy in
zero field \cite{Thomas1996,Thomas1998}. For our analysis here we
assume that the 0.7 anomaly is indeed related to such a spontaneous
spin splitting for the first subband. In high fields, all 12 QPCs
showed also for $N=2$ and higher a pronounced spin splitting into
two transconductance peaks, but these subbands did not clearly show
a zero-field splitting.

\begin{figure}[h!]
  \includegraphics[width=1\columnwidth]{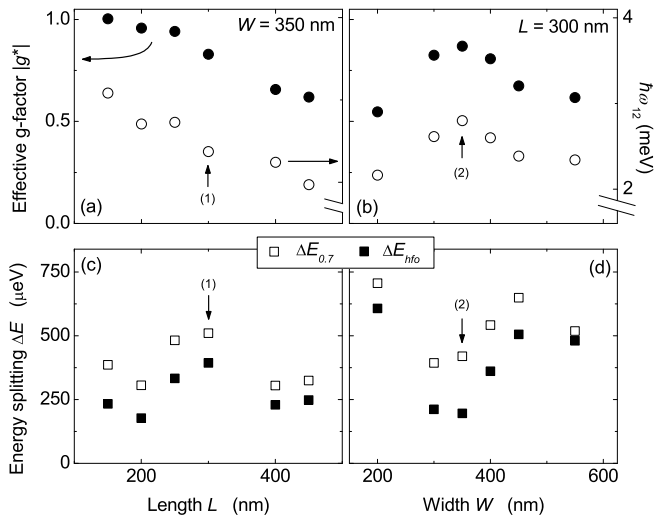}\\
  \caption{{\bf Geometry dependence of g-factor and energy splittings.
  a, b,} Effective g-factor $|g^*|$ (solid symbols, left axis) and subband spacing
  $\hbar \omega_{12}$ (open symbols, right axis) as a function of QPC
  length $L$ (with fixed width $W$ = 350 nm), and as a function QPC width $W$ (with
  fixed length $L$ = 300 nm). All data points are for the $N=1$ subband from results measured
  at 200~mK.
  The effective g-factor $|g^*|$ is enhanced as compared to the bulk 2DEG value
  (up to a factor $\sim 3$) and shows a clear
  dependence on $L$ and $W$. As a function of $L$ at fixed $W$ the enhancement is
  strongest for the shortest QPC.
  As a function of $W$, there is a maximum around $W = 350 \; {\rm nm}$.
  The subband spacing $\hbar \omega_{12}$ depends in a remarkably similar way on $L$ and
  $W$ as $|g^*|$. {\bf c, d,} The 0.7 energy splitting $\Delta E_{0.7}$ and high-field offset $\Delta E_{hfo}$ for the $N=1$ subband as a
  function of $L$ and $W$, derived from the same transconductance data as used for ({\bf a}) and ({\bf b}). The values of
  $\Delta E_{0.7}$ and $\Delta E_{hfo}$ vary with $L$ and $W$ in a irregular manner,
  but there is a strong correlation between $\Delta E_{0.7}$ and $\Delta E_{hfo}$.
  Data points labeled with (1) and (2) are from two different devices with identical values for $L$ and $W$.}
 \label{FIG:GfactorAndEnergySplittingVSShape}
\end{figure}

We studied how the spin splittings $\Delta E$ for $N=1,2,3$ increase
with magnetic field from $B=0\; {\rm T}$ up to 9~T
(Fig.~\ref{FIG:EnergySplittingVSMagneticField}b,c). We first
concentrate on data for $N=1$. At zero field $\Delta E$ shows the
splitting associated with the 0.7 anomaly, that we label $\Delta
E_{0.7}$. It is observed in all our QPCs with a typical value of
0.5~meV. At high fields $\Delta E$ has a linear slope similar to the
Zeeman effect. However, linear extrapolation of this slope down to
$B = 0$ shows that there is a large positive offset (unlike the
usual Zeeman effect). We characterize the slope with an effective
g-factor $|g^*| =\frac{1}{\mu_{B}} \frac{{\rm d}\Delta E}{{\rm d}B}$
(note that one should be careful to interpret $|g^*|$ as an absolute
indication for the g-factor of electrons in a QPC, since different
methods for extracting a g-factor can give different results
\cite{Cronenwett2002,Cronenwett2001}). The high-field offset from a
linear Zeeman effect is characterized with a parameter $\Delta
E_{hfo}$. Qualitatively, this type of data for $\Delta E$ looks
similar for all 12 QPCs, and we use a suitable fitting procedure
(see Methods) to characterize the traces for $N=1$ with the
parameters $\Delta E_{0.7}$, $\Delta E_{hfo}$ and $|g^*|$. Notably,
two-parameter fitting using spin-$\frac{1}{2}$ energy eigenvalues
with $\Delta E = \sqrt{ (\Delta E_{0.7})^2 +( |g^*| \mu_B B )^2}$
does not yield good fits. For the traces as in
Fig.~\ref{FIG:EnergySplittingVSMagneticField}b,c for $N=2,3$, we
cannot resolve a spin splitting at low fields, and we only provide
the parameters $\Delta E_{hfo}$ and $|g^*|$.

Fig.~\ref{FIG:EnergySplittingVSMagneticField}b,c show that $\Delta
E$ appears in high fields as the sum of the Zeeman effect and the
constant contribution $\Delta E_{hfo}$. This suggest that the
splittings in high field have, in particular for $N=1$, a
significant contribution from a field-independent exchange effect
that results from each subband being in a ferromagnetic
spin-polarized state. In high fields such an interpretation is less
ambiguous than for zero field (where the possibility of a
ferromagnetic ground state for spin-polarized subbands is the topic
of debate \cite{Klironomos2006,Jaksch2006}) since the Zeeman effect
suppresses spin fluctuations. Thus, measuring $\Delta E_{hfo}$ can
be used to determine this exchange splitting.

Fig.~\ref{FIG:GfactorAndEnergySplittingVSShape}a,b present how the
effective g-factor $|g^*|$ for $N=1$ varies with $L$ and $W$ of the
QPCs. It is strongly enhanced up to a factor $\sim 3$ with respect
to the g-factor for bulk 2DEG material \cite{Hannak1995}. This has
been observed before \cite{Thomas1998} and is also attributed to
many-body effects. In the same plot we present the subband spacing
$\hbar \omega_{12}$. The variation of $\hbar \omega_{12}$ is in good
agreement with an electrostatic analysis \cite{Davies1995} of the
degree of transverse confinement in the saddle-point-like potential
of the QPC. Briefly, short and narrow QPCs yield the strongest
transverse confinement
(Fig.~\ref{FIG:GfactorAndEnergySplittingVSShape}a). This is valid
down to the point where the QPC width $W \lesssim 2d$ (where $d$ the
depth of the 2DEG below the wafer surface), which results in the
maximum for $W = 350 \; {\rm nm}$ in
Fig.~\ref{FIG:GfactorAndEnergySplittingVSShape}b. Notably, the
values of $|g^*|$ and $\hbar \omega_{12}$ are clearly correlated.
While this is interesting on itself, for the current Letter we
restrict ourselves to the conclusion that the variation of these
effects with $L$ and $W$ is very regular. This indicates that part
of the physics of our set of 12 QPCs depends in a regular manner on
$L$ and $W$. This is also supported by the observation that the gate
voltage $V_g$ needed for pinching off a QPC varies in the expected
way with $L$ and $W$.

Next, we analyze how $\Delta E_{hfo}$ and $\Delta E_{0.7}$ depend on
$L$ and $W$. The open squares in
Fig.~\ref{FIG:GfactorAndEnergySplittingVSShape}c,d present this for
$\Delta E_{0.7}$. Overall, the dependence here is not very regular,
possibly indicating that the exact appearance of the otherwise
robust 0.7 anomaly is sensitive to small irregularities in the
potential that defines the QPC (only the data in
Fig.~\ref{FIG:GfactorAndEnergySplittingVSShape}d suggests an
anti-correlation with $\hbar \omega_{12}$). The black squares
present how $\Delta E_{hfo}$ for $N=1$ varies with $L$ and $W$. Also
here the dependence is irregular. Remarkably, the irregular
variations of $\Delta E_{0.7}$ and $\Delta E_{hfo}$ are clearly
correlated throughout our set of 12 QPCs. This means that $\Delta
E_{0.7}$, which is derived from data in zero field, is correlated
with $\Delta E_{hfo}$, which is derived from data taken at fields in
excess of 5~T. Furthermore, $\Delta E_{0.7}$ and $\Delta E_{hfo}$
have a similar order of magnitude. This points to the conclusion
that the spontaneous energy splitting of the 0.7 anomaly is
dominated by the same effect that causes the high-field offset
$\Delta E_{hfo}$. As we discussed, this is probably an exchange
contribution.

\begin{figure}[h!]
  \includegraphics[width=0.9\columnwidth]{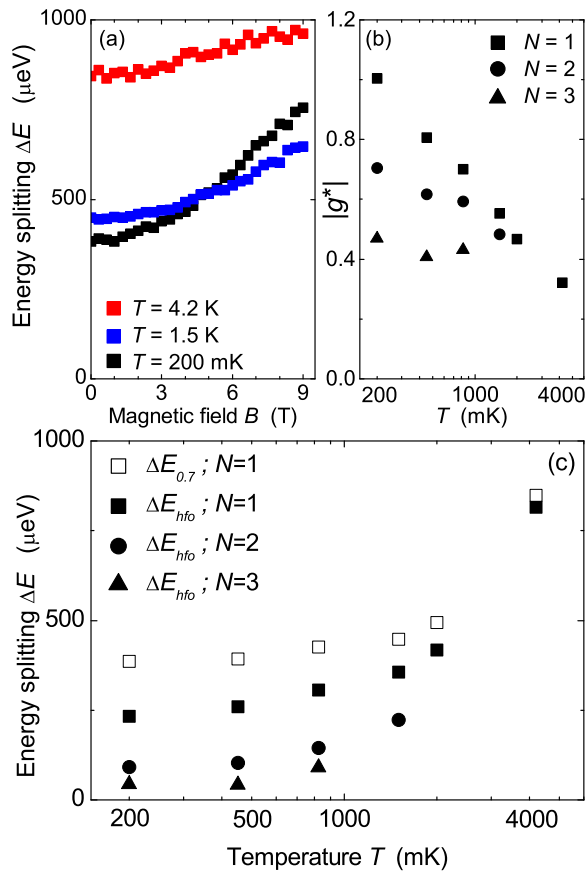}\\
  \caption{{\bf Temperature dependence of g-factor and energy splittings.
  a,} Energy splitting for $N=1$ as a function of magnetic field for
  different temperatures $T$, for a QPC with $L$ = 150 nm and $W$ = 350
  nm. {\bf b,} Effective g-factor $|g^*|$ as a function of temperature for the same QPC. The
  g-factor enhancement is strongest for the $N=1$ subband at the
  lowest temperature. For the $N=2$ and $N=3$ subband the g-factor
  is also enhanced at low temperatures. As the temperature is
  increased the g-factor enhancement is weaker for all subbands. {\bf c,} The 0.7 energy
  splitting $\Delta E_{0.7}$ and high-field offset $\Delta E_{hfo}$ as a function of temperature. The value for $\Delta
  E_{hfo}$ is highest for the $N=1$ subband and decreases to zero with increasing
  subband number. As the
  temperature is increased, the $\Delta E_{0.7}$ value as well as the $\Delta E_{hfo}$ values for $N=1,2,3$ strongly increase.
  The correlation between $\Delta E_{0.7}$ and $\Delta E_{hfo}$ remains present upon increasing the temperature.}
  \label{FIG:EnergySplittingVSTemperature}
\end{figure}

Using the temperature dependence of $\Delta E$ data
(Fig.~\ref{FIG:EnergySplittingVSTemperature}a), we find that the
correlation between $\Delta E_{0.7}$ and $\Delta E_{hfo}$ remains
intact at higher temperatures
(Fig.~\ref{FIG:EnergySplittingVSTemperature}c).
Figure~\ref{FIG:EnergySplittingVSTemperature}b shows that $|g^*|$
has a very different temperature dependence. This indicates that the
g-factor enhancement and the 0.7 anomaly arise from different
many-body effects. Figs.~\ref{FIG:EnergySplittingVSMagneticField}
and \ref{FIG:EnergySplittingVSTemperature} also present data for the
$N=2$ and $N=3$ subband. We analyzed the data for $N=2,3$ in the
very same way as for $N=1$, and the most important observation is
that the $\Delta E_{hfo}$ parameter for $N=2,3$ is much smaller than
for $N=1$, and often close to zero. A high $\Delta E_{hfo}$ value is
only observed for $N=1$, just as 0.7 anomaly itself. Finally, we
remark that most of our QPCs showed a zero-bias conductance peak in
non-linear transport near the 0.7 anomaly. The temperature and
magnetic field dependence of this peak (data not shown) is
consistent with reports \cite{Cronenwett2002} that relate the 0.7
anomaly to transport through a Kondo impurity.


Our main conclusion is that the many-body electron physics that
causes the spontaneous energy splitting of the 0.7 anomaly is
related to a field-independent exchange effect that contributes to
spin splittings in high magnetic fields. Our results are important
for theory work that aims at developing a consistent picture of
many-body effects in QPCs, and its consequences for transport of
spin-polarized electrons and spin coherence in nanodevices.


\vskip 4mm

\noindent {\bf METHODS} \vskip 2mm


{\bf Device fabrication}

Our devices were fabricated using a ${\rm GaAs}/{\rm Al}_{0.32}{\rm
Ga}_{0.68}{\rm As}$ heterostructure with a 2DEG at 114 nm below the
surface from modulation doping with Si. All QPCs were fabricated in
close proximity of each other on the same wafer to ensure the same
heterostructure properties for all QPCs. At 4.2 K, the mobility of
the 2DEG was $\mu = 159 \; {\rm m^{2}/Vs }$, and the electron
density $n_{s} = 2.14 \cdot 10^{15} \; {\rm m^{-2} }$. Depletion
gates were defined with standard electron-beam lithography and
lift-off techniques, using deposition of 15 nm of Au with a Ti
sticking layer. The reservoirs were connected to macroscopic leads
via Ohmic contacts, which were realized by annealing a thin Au/Ge/Ni
layer that was deposited on the surface.

{\bf Measurement techniques}

Measurements were performed in a dilution refrigerator with the
sample at temperatures from $\sim 5 \; {\rm mK}$ to 4.2~K. For all
our data the temperature dependence saturated when cooling below
$\sim 200$~mK. We therefore assume for this report that this is the
lowest effective electron temperature that could be achieved.

For measuring the differential conductance $G$ we used standard
lock-in techniques at 380 Hz, with an ac voltage bias $V_{ac}=10 \;
{\rm \mu V}$. Only the $V_{-}$ contact was connected to the grounded
shielding of our setup, and all gate voltages were applied with
respect to this ground. The in-plane magnetic field was applied
perpendicular to the current direction, and the current in the QPCs
was along the $[110]$ crystal orientation. Alignment of the sample
with the magnetic field was within 1$^\circ$, as determined from
Hall voltage measurements on the 2DEG.

We have data from 12 different QPCs from a set of 16 that we cooled
down. From these QPCs 4 could not be measured. For two this was due
to the presence of strong telegraph noise in conductance signals.
Two other QPCs did not show clear conductance plateaus.

{\bf Data analysis and fitting procedures}

For analyzing QPC conductance values we had to subtract a magnetic
field and temperature dependent series resistance (from the wiring
and filters, Ohmic contacts and 2DEG) from the transport data that
was obtained with a voltage-bias approach. The criterium here was to
make the observed conductance plateaus coincide with integer
multiples of $2e^{2}/h$ or $e^{2}/h$.

To determine the peak spacings in the transconductance traces, we
assume that each subband contributes in a parallel manner to the QPC
conductance. The transconductance curves can then be analyzed as a
superposition of peaks, with one (two) peak(s) per (spin-split)
subband. We carried out a least squares fitting with peak shapes
(for example Gaussian or Lorentzian, we checked that our results do
not depend significantly on the choice of peak shape we used here).
The specific shape of a step between the quantized conductance
plateaus depends on the shape of the saddle-point-like potential
that defines the QPC \cite{Buttiker1990}, and can result in
asymmetric transconductance peaks. We checked that this is not a
significant effect for our analysis. In the analysis of $\hbar
\omega_{12}$ and conversion of $\Delta V_{g}$ into $\Delta E$ we
observed a weak dependence on magnetic field and temperature, and
took this in account.

The energy splittings versus magnetic field from data taken at 200
mK (as in Figs.~\ref{FIG:EnergySplittingVSMagneticField}b,c) showed
for all 12 different QPCs a clear constant part (typically for
fields below 2~T), and a clear linear part (with an offset,
typically for fields above 5~T). On these parts of the data we
performed a least-square fitting with a constant or linear function
to determine the parameters $\Delta E_{0.7}$, $\Delta E_{hfo}$ and
$|g^*|$. We checked that the values of the derived parameters did
not significantly depend on the choice for these magnetic field
intervals that we used here.


\noindent {\bf Acknowledgements}

We thank B.~H.~J.~Wolfs, R.~N.~Schouten, S.~F.~Fischer,
C.~W.~J.~Beenakker and L.~S.~Levitov for help and useful
discussions, and the Dutch Foundation for Fundamental Research on
Matter (FOM), the Netherlands Organization for Scientific Research
(NWO), and the German BMBF (in the framework of the
nanoQUIT-program) for financial support.



\end{document}